# Numerical solutions of 2-D steady incompressible flow in a driven skewed cavity.

**Ercan Erturk [1], Bahtiyar Dursun**

Gebze Institute of Technology, Energy Systems Engineering Department, Gebze, Kocaeli 41400, Turkey

**Abstract**

The benchmark test case for non-orthogonal grid mesh, the "driven skewed cavity flow", first introduced by Demirdžić *et al.* [5] for skew angles of $\alpha = 30°$ and $\alpha = 45°$, is reintroduced with a more variety of skew angles. The benchmark problem has non-orthogonal, skewed grid mesh with skew angle ($\alpha$). The governing 2-D steady incompressible Navier-Stokes equations in general curvilinear coordinates are solved for the solution of driven skewed cavity flow with non-orthogonal grid mesh using a numerical method which is efficient and stable even at extreme skew angles. Highly accurate numerical solutions of the driven skewed cavity flow, solved using a fine grid ($512 \times 512$) mesh, are presented for Reynolds number of 100 and 1000 for skew angles ranging between $15° \leq \alpha \leq 165°$.

## 1. Introduction

In the literature, it is possible to find many numerical methods proposed for the solution of the steady incompressible N-S equations. These numerical methods are often tested on several benchmark test cases in terms of their stability, accuracy as well as efficiency. Among several benchmark test cases for steady incompressible flow solvers, the driven cavity flow is a very well known and commonly used benchmark problem. The reason why the driven cavity flow is so popular may be the simplicity of the geometry. In this flow problem, when the flow variables are nondimensionalized with the cavity length and the velocity of the lid, Reynolds number appears in the equations as an important flow parameter. Even though the geometry is simple and easy to apply in programming point of view, the cavity flow has all essential flow physics with counter rotating recirculating regions at the corners of the cavity. Among numerous papers found in the literature, Erturk *et al.* [6], Botella and Peyret [4], Schreiber and Keller [21], Li *et al.* [12], Wright and Gaskel [30], Erturk and Gokcol [7], Benjamin and Denny [2] and Nishida and Satofuka [16] are examples of numerical studies on the driven cavity flow.

Due to its simple geometry, the cavity flow is best solved in Cartesian coordinates with Cartesian grid mesh. Most of the benchmark test cases found in the literature have orthogonal geometries therefore they are best solved with orthogonal grid mesh. However often times the real life flow problems have much more complex geometries than that of the driven cavity flow. In most cases, researchers have to deal with non-orthogonal geometries with non-orthogonal grid mesh. In a non-orthogonal grid mesh, when the governing equations are formulated in general curvilinear coordinates, cross derivative terms appear in the equations. Depending on the skewness of the grid mesh, these cross derivative terms can be very significant and can affect the numerical stability as well as the accuracy of the numerical method used for the solution. Even though, the driven cavity flow benchmark problem serves for comparison between numerical methods, the flow is far from simulating the real life fluid problems with complex geometries with non-orthogonal grid mesh. The numerical performances of numerical methods on orthogonal grids may or may not be the same on non-orthogonal grids.

Unfortunately, there are not much benchmark problems with non-orthogonal grids for numerical methods to compare solutions with each other. Demirdžić *et al.* [5] have introduced the driven skewed cavity flow as a test case for non-orthogonal grids. The test case is similar to driven cavity flow but the geometry is a parallelogram rather than a square. In this test case, the skewness of the geometry can be easily changed by changing the skew angle ($\alpha$). The skewed cavity problem is a perfect test case for body fitted non-orthogonal grids and yet it is as simple as the cavity flow in terms of programming point of view. Later Oosterlee et al. [17], Louaked et al. [13], Roychowdhury et al. [20], Xu and Zhang [31], Wang and Komori [28], Xu and Zhang [32], Tucker and Pan [27], Brakkee et al. [3], Pacheco and Peck [18], Teigland and Eliassen [25], Lai and Yan [11] and Shklyar and Arbel [22] have solved the same benchmark problem. In all these studies, the solution of the driven skewed cavity flow is presented for Reynolds numbers of 100 and 1000 for only two different skew angles which are $\alpha = 30°$ and $\alpha = 45°$ and also the maximum number of grids used in these studies is $320 \times 320$.

---
[1]Corresponding author, e-mail: ercanerturk@gyte.edu.tr, URL: http://www.cavityflow.com

Perić [19] considered the 2-D flow in a skewed cavity and he stated that the governing equations fail to converge for $\alpha < 30°$. The main motivation of this study is then to reintroduce the skewed cavity flow problem with a wide range of skew angle ($15° \leq \alpha \leq 165°$) and present detailed tabulated results obtained using a fine grid mesh with $512 \times 512$ points for future references.

Erturk et al. [6] have introduced an efficient, fast and stable numerical formulation for the steady incompressible Navier-Stokes equations. Their methods solve the streamfunction and vorticity equations separately, and the numerical solution of each equation requires the solution of two tridiagonal systems. Solving tridiagonal systems are computationally efficient and therefore they were able to use very fine grid mesh in their solution. Using this numerical formulation, they have solved the very well known benchmark problem, the steady flow in a square driven cavity, up to Reynolds number of 21000 using a $601 \times 601$ fine grid mesh. Their formulation proved to be stable and effective at very high Reynolds numbers ([6], [7], [8]).

In this study, the numerical formulation introduced by Erturk et al. [6] will be applied to Navier-Stokes equations in general curvilinear coordinates and the numerical solutions of the driven skewed cavity flow problem with body fitted non-orthogonal skewed grid mesh will be presented. By considering a wide range of skew angles, the efficiency of the numerical method will be tested for grid skewness especially at extreme skew angles. The numerical solutions of the flow in a skewed cavity will be presented for Reynolds number of 100 and 1000 for a wide variety of skew angles ranging between $\alpha = 15°$ and $\alpha = 165°$ with $\Delta\alpha = 15°$ increments.

## 2. Numerical Formulation

For two-dimensional and axi-symmetric flows it is convenient to use the streamfunction ($\psi$) and vorticity ($\omega$) formulation of the Navier-Stokes equations. In non-dimensional form, they are given as

$$\psi_{xx} + \psi_{yy} = -\omega \tag{1}$$

$$\psi_y \omega_x - \psi_x \omega_y = \frac{1}{Re}(\omega_{xx} + \omega_{yy}) \tag{2}$$

where, $Re$ is the Reynolds number, and $x$ and $y$ are the Cartesian coordinates. We consider the governing Navier-Stokes equations in general curvilinear coordinates as the following

$$(\xi_x^2 + \xi_y^2)\psi_{\xi\xi} + (\eta_x^2 + \eta_y^2)\psi_{\eta\eta} + (\xi_{xx} + \xi_{yy})\psi_\xi + (\eta_{xx} + \eta_{yy})\psi_\eta + 2(\xi_x\eta_x + \xi_y\eta_y)\psi_{\xi\eta} = -\omega \tag{3}$$

$$(\xi_x\eta_y)\psi_\eta\omega_\xi - (\xi_x\eta_y)\psi_\xi\omega_\eta = \frac{1}{Re}((\xi_x^2 + \xi_y^2)\omega_{\xi\xi} + (\eta_x^2 + \eta_y^2)\omega_{\eta\eta}$$
$$+(\xi_{xx} + \xi_{yy})\omega_\xi + (\eta_{xx} + \eta_{yy})\omega_\eta + 2(\xi_x\eta_x + \xi_y\eta_y)\omega_{\xi\eta}) \tag{4}$$

Following Erturk *et al.* [6], first pseudo time derivatives are assigned to streamfunction and vorticity equations and using an implicit Euler time step for these pseudo time derivatives, the finite difference formulations in operator notation become the following

$$(1 - \Delta t(\xi_x^2 + \xi_y^2)\delta_{\xi\xi} - \Delta t(\eta_x^2 + \eta_y^2)\delta_{\eta\eta} - \Delta t(\xi_{xx} + \xi_{yy})\delta_\xi - \Delta t(\eta_{xx} + \eta_{yy})\delta_\eta)\psi^{n+1}$$
$$= \psi^n + 2\Delta t(\xi_x\eta_x + \xi_y\eta_y)\psi^n_{\xi\eta} + \Delta t\omega^n \qquad (5)$$

$$(1 - \frac{\Delta t}{Re}(\xi_x^2 + \xi_y^2)\delta_{\xi\xi} - \frac{\Delta t}{Re}(\eta_x^2 + \eta_y^2)\delta_{\eta\eta} - \frac{\Delta t}{Re}(\xi_{xx} + \xi_{yy})\delta_\xi - \frac{\Delta t}{Re}(\eta_{xx} + \eta_{yy})\delta_\eta$$
$$+ \Delta t(\xi_x\eta_y)\psi^n_\eta\delta_\xi - \Delta t(\xi_x\eta_y)\psi^n_\xi\delta_\eta)\omega^{n+1}$$
$$= \omega^n + \frac{2\Delta t}{Re}(\xi_x\eta_x + \xi_y\eta_y)\omega^n_{\xi\eta} \qquad (6)$$

Where $\delta_{\xi\xi}$ and $\delta_{\eta\eta}$ denote the second order finite difference operators, and similarly $\delta_\xi$ and $\delta_\eta$ denote the first order finite difference operators in $\xi$- and $\eta$-direction respectively. The equations above are in implicit form and require the solution of a large matrix at every pseudo time iteration which is computationally inefficient. Instead these equations are spatially factorized such that

$$(1 - \Delta t(\xi_x^2 + \xi_y^2)\delta_{\xi\xi} - \Delta t(\xi_{xx} + \xi_{yy})\delta_\xi)(1 - \Delta t(\eta_x^2 + \eta_y^2)\delta_{\eta\eta} - \Delta t(\eta_{xx} + \eta_{yy})\delta_\eta)\psi^{n+1}$$
$$= \psi^n + 2\Delta t(\xi_x\eta_x + \xi_y\eta_y)\psi^n_{\xi\eta} + \Delta t\omega^n \qquad (7)$$

$$(1 - \frac{\Delta t}{Re}(\xi_x^2 + \xi_y^2)\delta_{\xi\xi} - \frac{\Delta t}{Re}(\xi_{xx} + \xi_{yy})\delta_\xi + \Delta t(\xi_x\eta_y)\psi^n_\eta\delta_\xi)$$
$$\times (1 - \frac{\Delta t}{Re}(\eta_x^2 + \eta_y^2)\delta_{\eta\eta} - \frac{\Delta t}{Re}(\eta_{xx} + \eta_{yy})\delta_\eta - \Delta t(\xi_x\eta_y)\psi^n_\xi\delta_\eta)\omega^{n+1}$$
$$= \omega^n + \frac{2\Delta t}{Re}(\xi_x\eta_x + \xi_y\eta_y)\omega^n_{\xi\eta} \qquad (8)$$

The advantage of these equations are that each equation require the solution of a tridiagonal systems that can be solved very efficiently using the Thomas algorithm. It can be shown that approximate factorization introduces additional second order terms ($\mathcal{O}(\Delta t^2)$) in these equations. In order for the equations to have the correct physical representation, to cancel out the second order terms due to factorization the same terms are added to the right hand side of the equations. The reader is referred to Erturk *et al.* [6] for more details of the numerical method. The final form of the equations take the following form

$$(1 - \Delta t(\xi_x^2 + \xi_y^2)\delta_{\xi\xi} - \Delta t(\xi_{xx} + \xi_{yy})\delta_\xi)(1 - \Delta t(\eta_x^2 + \eta_y^2)\delta_{\eta\eta} - \Delta t(\eta_{xx} + \eta_{yy})\delta_\eta)\psi^{n+1}$$
$$= \psi^n + 2\Delta t(\xi_x\eta_x + \xi_y\eta_y)\psi^n_{\xi\eta} + \Delta t\omega^n$$
$$(\Delta t(\xi_x^2 + \xi_y^2)\delta_{\xi\xi} + \Delta t(\xi_{xx} + \xi_{yy})\delta_\xi)(\Delta t(\eta_x^2 + \eta_y^2)\delta_{\eta\eta} + \Delta t(\eta_{xx} + \eta_{yy})\delta_\eta)\psi^n \qquad (9)$$

$$(1 - \frac{\Delta t}{Re}(\xi_x^2 + \xi_y^2)\delta_{\xi\xi} - \frac{\Delta t}{Re}(\xi_{xx} + \xi_{yy})\delta_\xi + \Delta t(\xi_x\eta_y)\psi^n_\eta\delta_\xi)$$
$$\times (1 - \frac{\Delta t}{Re}(\eta_x^2 + \eta_y^2)\delta_{\eta\eta} - \frac{\Delta t}{Re}(\eta_{xx} + \eta_{yy})\delta_\eta - \Delta t(\xi_x\eta_y)\psi^n_\xi\delta_\eta)\omega^{n+1}$$
$$= \omega^n + \frac{2\Delta t}{Re}(\xi_x\eta_x + \xi_y\eta_y)\omega^n_{\xi\eta}$$

$$(\frac{\Delta t}{Re}(\xi_x^2 + \xi_y^2)\delta_{\xi\xi} + \frac{\Delta t}{Re}(\xi_{xx} + \xi_{yy})\delta_\xi - \Delta t(\xi_x \eta_y)\psi_\eta^n \delta_\xi)$$

$$\times (\frac{\Delta t}{Re}(\eta_x^2 + \eta_y^2)\delta_{\eta\eta} + \frac{\Delta t}{Re}(\eta_{xx} + \eta_{yy})\delta_\eta + \Delta t(\xi_x \eta_y)\psi_\xi^n \delta_\eta)\omega^n \qquad (10)$$

## 3. Driven Skewed Cavity Flow

Fig. 1 illustrates the schematic view of the benchmark problem, the driven skewed cavity flow. We will consider the most general case where the skew angle can be $\alpha > 90°$ or $\alpha < 90°$.

In order to calculate the metrics, the grids in the physical domain are mapped onto orthogonal grids in the computational grids as shown in Fig. 2. The inverse transformation metrics are calculated using central differences, as an example $\frac{\partial \Omega}{\partial u} = \frac{x_{i+1,j} - x_{i-1,j}}{2\Delta \xi} = \frac{2\frac{1}{N}}{2} = \frac{1}{N}$. Similarly, inverse transformation metrics are calculated as the following

$$x_\xi = \frac{1}{N} \quad , \quad x_\eta = \frac{\cos\alpha}{N} \quad , \quad y_\xi = 0 \quad , \quad y_\eta = \frac{\sin\alpha}{N} \qquad (11)$$

where $N$ is the number of grid points. We consider a ($N \times N$) grid mesh. The determinant of the Jacobian matrix is found as

$$|J| = x_\xi y_\eta - x_\eta y_\xi = \frac{\sin\alpha}{N^2} \qquad (12)$$

The transformation metrics are defined as

$$\xi_x = \frac{1}{|J|} y_\eta \quad , \quad \xi_y = \frac{-1}{|J|} x_\eta \quad , \quad \eta_x = \frac{-1}{|J|} y_\xi \quad , \quad \eta_y = \frac{1}{|J|} x_\xi \qquad (13)$$

Substituting Equations (11) and (12) into (13), the transformation metrics are obtained as the following

$$\xi_x = N \quad , \quad \xi_y = \frac{-N\cos\alpha}{\sin\alpha} \quad , \quad \eta_x = 0 \quad , \quad \eta_y = \frac{N}{\sin\alpha} \qquad (14)$$

Note that since we use equal grid spacing, the second order transformation metrics will be all equal to zero such that for example

$$\xi_{xx} = \frac{\partial(\xi_x)}{\partial x} = \xi_x \frac{\partial(\xi_x)}{\partial \xi} + \eta_x \frac{\partial(\xi_x)}{\partial \eta} = 0 \qquad (15)$$

Hence

$$\xi_{xx} = \xi_{yy} = \eta_{xx} = \eta_{yy} = 0 \qquad (16)$$

These calculated metrics are substituted into Equations (9) and (10) and the final form of the numerical equations become as the following

$$(1-\Delta t(\frac{N^2}{sin^2\alpha})\delta_{\xi\xi})(1-\Delta t(\frac{N^2}{sin^2\alpha})\delta_{\eta\eta})\psi^{n+1}$$
$$=\psi^n+2\Delta t(\frac{-N^2\cos\alpha}{sin^2\alpha})\psi^n_{\xi\eta}+\Delta t\omega^n+(\Delta t(\frac{N^2}{sin^2\alpha})\delta_{\xi\xi})(\Delta t(\frac{N^2}{sin^2\alpha})\delta_{\eta\eta})\psi^n \quad (17)$$

$$(1-\frac{\Delta t}{Re}(\frac{N^2}{sin^2\alpha})\delta_{\xi\xi}+\Delta t(\frac{N^2}{sin^2\alpha})\psi^n_\eta\delta_\xi)(1-\frac{\Delta t}{Re}(\frac{N^2}{sin^2\alpha})\delta_{\eta\eta}-\Delta t(\frac{N^2}{sin^2\alpha})\psi^n_\xi\delta_\eta)\omega^{n+1}$$
$$=\omega^n+\frac{2\Delta t}{Re}(\frac{-N^2\cos\alpha}{sin^2\alpha})\omega^n_{\xi\eta}$$
$$+(\frac{\Delta t}{Re}(\frac{N^2}{sin^2\alpha})\delta_{\xi\xi}-\Delta t(\frac{N^2}{sin\alpha})\psi^n_\eta\delta_\xi)(\frac{\Delta t}{Re}(\frac{N^2}{sin^2\alpha})\delta_{\eta\eta}+\Delta t(\frac{N^2}{sin\alpha})\psi^n_\xi\delta_\eta)\omega^n \quad (18)$$

The solution methodology of each of the above two equations, Equations (17) and (18), involves a two-stage time-level updating. First the streamfunction equation (17) is solved, and for this, the variable $f$ is introduced such that

$$(1-\Delta t(\frac{N^2}{sin^2\alpha})\delta_{\eta\eta})\psi^{n+1}=f \quad (19)$$

where

$$(1-\Delta t(\frac{N^2}{sin^2\alpha})\delta_{\xi\xi})f=\psi^n+2\Delta t(\frac{-N^2\cos\alpha}{sin^2\alpha})\psi^n_{\xi\eta}+\Delta t\omega^n$$
$$+(\Delta t(\frac{N^2}{sin^2\alpha})\delta_{\xi\xi})(\Delta t(\frac{N^2}{sin^2\alpha})\delta_{\eta\eta})\psi^n \quad (20)$$

In Equation (20) $f$ is the only unknown variable. First, this Equation (20) is solved for $f$ at each grid point. Following this, the streamfunction ($\psi$) variable is advanced into the new time level using Equation (19).

Then the vorticity equation (18) is solved, and in a similar fashion, the variable $g$ is introduced such that

$$(1-\frac{\Delta t}{Re}(\frac{N^2}{sin^2\alpha})\delta_{\eta\eta}-\Delta t(\frac{N^2}{sin\alpha})\psi^n_\xi\delta_\eta)\omega^{n+1}=g \quad (21)$$

where

$$(1-\frac{\Delta t}{Re}(\frac{N^2}{sin^2\alpha})\delta_{\xi\xi}+\Delta t(\frac{N^2}{sin\alpha})\psi^n_\eta\delta_\xi)g=\omega^n+\frac{2\Delta t}{Re}(\frac{-N^2\cos\alpha}{sin^2\alpha})\omega^n_{\xi\eta}$$
$$+(\frac{\Delta t}{Re}(\frac{N^2}{sin^2\alpha})\delta_{\xi\xi}-\Delta t(\frac{N^2}{sin\alpha})\psi^n_\eta\delta_\xi)(\frac{\Delta t}{Re}(\frac{N^2}{sin^2\alpha})\delta_{\eta\eta}+\Delta t(\frac{N^2}{sin\alpha})\psi^n_\xi\delta_\eta)\omega^n \quad (22)$$

As with $f$, first the variable $g$ is determined at every grid point using Equation (22), then vorticity ($\omega$) variable is advanced into the next time level using Equation (21).

## 3.1 Boundary Conditions

In the computational domain the velocity components are defined as the following

$$u = \psi_y = \xi_y \psi_\xi + \eta_y \psi_\eta = \frac{-N\cos\alpha}{\sin\alpha}\psi_\xi + \frac{N}{\sin\alpha}\psi_\eta \tag{23}$$

$$v = -\psi_x = -\xi_x \psi_\xi - \eta_x \psi_\eta = -N\psi_\xi \tag{24}$$

On the left wall boundary we have

$$\psi_{0,j} = 0 \quad , \quad \psi_\eta \big|_{0,j} = 0 \quad , \quad \psi_{\eta\eta}\big|_{0,j} = 0 \tag{25}$$

where the subscripts 0 and $j$ are the grid indexes. Also on the left wall, the velocity is zero ($u = 0$ and $v = 0$). Using Equations (23) and (24) we obtain

$$\psi_\xi \big|_{0,j} = 0 \tag{26}$$

and also

$$\psi_{\xi\eta} = \frac{\partial(\psi_\xi)}{\partial \eta}\bigg|_{0,j} = 0 \tag{27}$$

Therefore, substituting these into the streamfunction Equation (3) and using Thom's formula [26], on the left wall boundary the vorticity is calculated as the following

$$\omega_{0,j} = -\frac{2N^2 \psi_{1,j}}{\sin^2 \alpha} \tag{28}$$

Similarly the vorticity on the right wall ($\omega_{N,j}$) and the vorticity on the bottom wall ($\omega_{i,0}$) are defined as the following

$$\omega_{N,j} = -\frac{2N^2 \psi_{N-1,j}}{\sin^2 \alpha} \quad , \quad \omega_{i,0} = -\frac{2N^2 \psi_{i,1}}{\sin^2 \alpha} \tag{29}$$

On the top wall the $u$-velocity is equal to $u = 1$. Following the same procedure, the vorticity on the top wall is found as

$$\omega_{i,N} = -\frac{2N^2 \psi_{i,N-1}}{\sin^2 \alpha} - \frac{2N}{\sin\alpha} \tag{30}$$

We note that, it is well understood ([10], [15], [23], [29]) that, even though Thom's method is locally first order accurate, the global solution obtained using Thom's method preserves second order accuracy. Therefore in this study, since three point second order central difference is used inside the skewed cavity and Thom's method is used at the wall boundary conditions, the presented solutions are second order accurate.

In the skewed driven cavity flow, the corner points are singular points for vorticity. We note that due to the skew angle, the governing equations have cross derivative terms and because of these cross derivative terms the computational stencil includes 3×3 grid points. Therefore, the solution at the first diagonal grid points near the corners of the cavity require the vorticity values at the corner points. For square driven cavity flow Gupta *et al.* [9] have introduced an explicit asymptotic solution in the neighborhood of sharp corners. Similarly, Störtkuhl *et al.* [24] have presented an analytical asymptotic solutions near the corners of cavity and using finite element bilinear shape functions they also have presented a singularity removed boundary condition for vorticity at the corner points as well as at the wall points. In this study we follow Störtkuhl *et al.* [24] and use the following expression for calculating vorticity values at the corners of the skewed cavity

$$\frac{N^2}{3\sin^2\alpha}\begin{bmatrix} \bullet & \bullet & \bullet \\ \bullet & -2 & \frac{1}{2} \\ \bullet & \frac{1}{2} & 1 \end{bmatrix}\psi + \frac{1}{9}\begin{bmatrix} \bullet & \bullet & \bullet \\ \bullet & 1 & \frac{1}{2} \\ \bullet & \frac{1}{2} & \frac{1}{4} \end{bmatrix}\omega = -\frac{VN}{2\sin\alpha} \tag{31}$$

where $V$ is the speed of the wall which is equal to 1 for the upper two corners and it is equal to 0 for the bottom two corners. The reader is referred to Störtkuhl *et al.* [24] for details.

## 4. Results

The steady incompressible flow in a driven skewed cavity is numerically solved using the described numerical formulation and boundary conditions. We have considered two Reynolds numbers, *Re*=100 and *Re*=1000. For these two Reynolds numbers we have varied the skew-angle ($\alpha$) from $\alpha = 15°$ $\alpha = 165°$ with $\Delta\alpha = 15°$ increments. We have solved the introduced problem with a $512\times512$ grid mesh, for the two Reynolds number and for all the skew angles considered.

During the iterations as a measure of the convergence to the steady state solution, we monitored three error parameters. The first error parameter, *ERR1*, is defined as the maximum absolute residual of the finite difference equations of the steady streamfunction and vorticity equations in general curvilinear coordinates, Equations (3) and (4). These are respectively given as

$$ERR1_\psi = \max\left(\text{abs}\left(\left|\frac{N^2}{\sin^2 a}\psi_{\xi\xi} + \frac{N^2}{\sin^2 a}\psi_{\eta\eta} - 2\frac{N^2\cos\alpha}{\sin^2 a}\psi_{\xi\eta} + \omega\right|^n_{i,j}\right)\right)$$

$$ERR1_w = \max\left(\text{abs}\left(\left|\frac{1}{Re}\frac{N^2}{\sin^2 a}\omega_{\xi\xi} + \frac{1}{Re}\frac{N^2}{\sin^2 a}\omega_{\eta\eta} - \frac{N^2}{\sin^2 a}\psi_\eta\omega_\xi + \frac{N^2}{\sin^2 a}\psi_\xi\omega_\eta - \frac{2}{Re}\frac{N^2\cos\alpha}{\sin^2 a}\omega_{\xi\eta}\right|^n_{i,j}\right)\right) \tag{32}$$

The magnitude of *ERR1* is an indication of the degree to which the solution has converged to steady state. In the limit *ERR1* would be zero.

The second error parameter, *ERR2*, is defined as the maximum absolute difference between an iteration time step in the streamfunction and vorticity variables. These are respectively given as

$$ERR2_\psi = \max\left(\text{abs}\left(\psi^{n+1}_{i,j} - \psi^n_{i,j}\right)\right)$$
$$ERR2_\omega = \max\left(\text{abs}\left(\omega^{n+1}_{i,j} - \omega^n_{i,j}\right)\right) \tag{33}$$

$ERR2$ gives an indication of the significant digit of the streamfunction and vorticity variables are changing between two time levels.

The third error parameter, $ERR3$, is similar to $ERR2$, except that it is normalized by the representative value at the previous time step. This then provides an indication of the maximum percent change in $\psi$ and $\omega$ at each iteration step. $ERR3$ is defined as

$$ERR3_\psi = \max\left(\text{abs}\left(\frac{\psi_{i,j}^{n+1} - \psi_{i,j}^n}{\psi_{i,j}^n}\right)\right)$$

$$ERR3_\omega = \max\left(\text{abs}\left(\frac{\omega_{i,j}^{n+1} - \omega_{i,j}^n}{\omega_{i,j}^n}\right)\right) \qquad (34)$$

In our computations, for every Reynolds numbers and for every skew angles, we considered that convergence was achieved when both $ERR1_\psi$ and $ERR1_\omega$ were less than $10^{-10}$. Such a low value was chosen to ensure the accuracy of the solution. At these residual levels, the maximum absolute difference in streamfunction value between two time steps, $ERR2_\psi$, was in the order of $10^{-17}$ and for vorticity, $ERR2_\omega$, it was in the order of $10^{-15}$. And also at these convergence levels, between two time steps the maximum absolute normalized difference in streamfunction, $ERR3_\psi$, and in vorticity, $ERR3_\omega$, was in the order of $10^{-14}$, and $10^{-13}$ respectively.

We note that at extreme skew angles, convergence to such low residuals is necessary. For example, at skew angle $\alpha = 15°$ at the bottom left corner, and at skew angle $\alpha = 165°$ at the bottom right corner, there appears progressively smaller counter rotating recirculating regions in accordance with Moffatt [14]. In these recirculating regions confined in the sharp corner, the value of streamfunction variable is getting extremely smaller as the size of the recirculating region gets smaller towards the corner. Therefore, it is crucial to have convergence to such low residuals especially at extreme skew angles.

Before solving the skewed driven cavity flow at different skew angles first we have solved the square driven cavity flow to test the accuracy of the solution. The square cavity is actually a special case for skewed cavity and obtained when the skew angle is chosen as $\alpha = 90°$. For the square driven cavity flow, the streamfunction and the vorticity values at the center of the primary vortex and the location of this center are tabulated in Table 1 for Reynolds numbers of $Re$=1000, together with results found in the literature. The present results are almost exactly the same with that of Erturk *et al.* [6]. This was expected since in both studies the same number of grid points were used and also the spatial accuracy of both the boundary condition approximations and the solutions were the same. Furthermore the presented results are in very good agreement with that of highly accurate spectral solutions of Botella and Peyret [4] and extrapolated solutions of Schreiber and Keller [21] and also fourth order solutions of Erturk and Gokcol [7] with approximately less than 0.18% and 0.14% difference in streamfunction and vorticity variables respectively. For all the skew angles considered in this study ($15° \leq \alpha \leq 165°$) we expect to have the same level of accuracy we achieved forc. With Li *et al.* [12], Wright and Gaskel [30], Benjamin and Denny [2] and Nishida and Satofuka [16] again our solutions compare good.

After validating our solution for $\alpha = 90°$, we decided to validate our solutions at different skew angles. In order to do this we compare our results with the results found in the literature. At this point, we would like to note that in the literature among the studies that have solved the skewed cavity flow ([5], [17], [13], [20], [31], [28], [32], [27], [3], [18], [25], [11] and [22]), only Demirdžić *et al.* [5], Oosterlee *et al.* [17], Shklyar and Arbel [22] and Louaked *et al.* [13] have presented tabulated results therefore we will mainly compare our results with those studies.

As mentioned earlier, Demirdžić et al. [5] have presented solutions for skewed cavity for Reynolds number of 100 and 1000 for skewed angles of $\alpha = 45°$ and $\alpha = 30°$. Figure 3 compares our results of *u*-velocity along line A-B and *v*-velocity along line C-D with that of Demirdžić et al. [5] for *Re*=100 and 1000 for $\alpha = 45°$, and also Figure 4 compares the same for $\alpha = 30°$. Our results agree excellent with results of Demirdžić et al. [5].

Table 2 compares our results of the minimum and also maximum streamfunction value and also their location for Reynolds numbers of 100 and 1000 for skew angles of $\alpha = 30°$ and $\alpha = 45°$ with results of Demirdžić et al. [5], Oosterlee et al. [17], Louaked et al. [13] and Shklyar and Arbel [22]. The results of this study and the results of Demirdžić et al. [5] and also those of Oosterlee et al. [17], Shklyar and Arbel [22] and Louaked et al. [13] agree well with each other, although we believe that our results are more accurate since in this study a very fine grid mesh is used.

Figure 5 to Figure 8 show the streamline and also vorticity contours for *Re*=100 and *Re*=1000 for skew angles from $\alpha = 15°$ to $\alpha = 165°$ with $\Delta\alpha = 15°$ increments. As it is seen from these contour figures of streamfunction and vorticity, the solutions obtained are very smooth without any wiggles in the contours even at extreme skew angles.

We have solved the incompressible flow in a skewed driven cavity numerically and compared our numerical solution with the solutions found in the literature for $\alpha = 90°$, 30° and 45°, and good agreement is found. We, then, have presented solutions for $15° \leq \alpha \leq 165°$. Since we could not find solutions in the literature to compare with our presented solutions other than $\alpha = 90°$, 30° and 45°, in order to demonstrate the accuracy of the numerical solutions we presented, a good mathematical check would be to check the continuity of the fluid, as suggested by Aydin and Fenner [1]. We have integrated the *u*-velocity and *v*-velocity profiles along line A-B and line C-D, passing through the geometric center of the cavity shown by the red dotted line in Figure 1, in order to obtain the net volumetric flow rate through these sections. Through section A-B, the volumetric flow rate is $Q_{AB} = \left| \int u \, dy + \int v \, dx \right|$ and through section C-D it is $Q_{CD} = \left| \int v \, dx \right|$. Since the flow is incompressible, the net volumetric flow rate passing through these sections should be equal to zero, $Q = 0$. Using Simpson's rule for the integration, the volumetric flow rates $Q_{AB}$ and $Q_{CD}$ are calculated for every skew angle ($\alpha$) and every Reynolds number considered. In order to help quantify the errors, the obtained volumetric flow rate values are normalized by the absolute total flow rate through the corresponding section at the considered *Re* and $\alpha$. Hence $Q_{AB}$ is normalized by $\int |u| \, dy + \int |v| \, dx$ and similarly $Q_{CD}$ is normalized by $Q_{CD} = \int |v| \, dx$. Table 3 tabulates the normalized volumetric flow rates through the considered sections. We note that in an integration process the numerical errors will add up, nevertheless, the normalized volumetric flow rate values tabulated in Table 3 are close to zero, such that they can be considered as $Q_{AB} \approx Q_{CD} \approx 0$. This mathematical check on the conservation of the continuity shows that our numerical solution is indeed very accurate at the considered skew angles and Reynolds numbers.

We note that, to the authors best knowledge, in the literature there is not a study that considered the skewed cavity flow at the skew angles used in the present study other than $\alpha = 30°$ and $\alpha = 45°$. The solutions presented in this study are unique therefore, for future references, in Table 4 we have tabulated the minimum and also maximum streamfunction values and their locations and also the vorticity value at these points for Reynolds number of 100 and 1000 for all the skew angles considered, from $\alpha = 15°$ to $\alpha = 165°$ with $\Delta\alpha = 15°$ increments. In this table the interesting point is, at Reynolds number of 1000, the strength of vorticity (absolute value of the vorticity) at the center of the primary vortex decrease as the skew angle increase from $\alpha = 15°$ to $\alpha = 90°$, having the minimum value at $\alpha = 90°$. As the skew angle increase further from $\alpha = 90°$ to $\alpha = 165°$, the strength of vorticity at the center of the primary vortex increase. At this Reynolds number, *Re*=1000, the streamfunction value at the center of the primary vortex also show the same type of behavior, where the value of the streamfunction start to increase as the skew angle increase until $\alpha = 90°$, then start to decrease as the skew angle increase further. However at Reynolds number of 100, the minimum value of the strength of vorticity and also the maximum value of the streamfunction at the center of the primary vortex occur at $\alpha = 105°$. In order to explain this behavior, we decided to look at the location of the center of the primary vortex at different skew angles. The strength of vorticity at the center of the primary vortex is proportional with the vertical distance between the center of the primary vortex and the top moving lid. As the center of the primary vortex move away from the top moving wall, the strength of vorticity at the center should decrease. Figure 9 shows the vertical distance between the center of the primary vortex and the top moving lid and as a function of the skew angle for both *Re*=100 and 1000. As this figure show at *Re*=100 the eye of the primary vortex is at its farthest position from the top moving lid at $\alpha = 105°$ where as at *Re*=1000 the same occurs at $\alpha = 90°$, explaining the minimum vorticity strength we obtain for the primary vortex at $\alpha = 105°$ for *Re*=100 and also at $\alpha = 90°$ for *Re*=1000. For future references, in Table 5 and 6 we have tabulated the *u*-velocity profiles along line A-B (shown in Figure 1) for Reynolds number of 100 and 1000 respectively, and similarly, in Table 7 and 8 we have tabulated the *v*-velocity profiles along line C-D (also shown in Figure 1) for Reynolds number of 100 and 1000 respectively.

## 5. Conclusions

In this study the benchmark test case for non-orthogonal grid mesh, the skewed cavity flow introduced by Demirdžić *et al.* [5] for skew angles of $\alpha = 30°$ and $\alpha = 45°$, is reintroduced with a variety of skew angles. The skewed cavity flow is considered for skew angles ranging between $15° \leq \alpha \leq 165°$ with $\Delta\alpha = 15°$ increments, for *Re*=100 and *Re*=1000. The governing Navier-Stokes equations are considered in most general form, in general curvilinear coordinates. The non-orthogonal grids are mapped onto a computational domain. Using the numerical formulation introduced by Erturk *et al.* [6], fine grid solutions of streamfunction and vorticity equations are obtained with very low residuals. The numerical formulation of Erturk *et al.* [6] have proved to be very effective on non-orthogonal problems with non-orthogonal grid mesh even at extreme skew angles. The driven skewed cavity flow problem is a challenging problem and it can be a perfect benchmark test case for numerical methods to test performances on non-orthogonal grid meshes. For future references detailed results are tabulated.

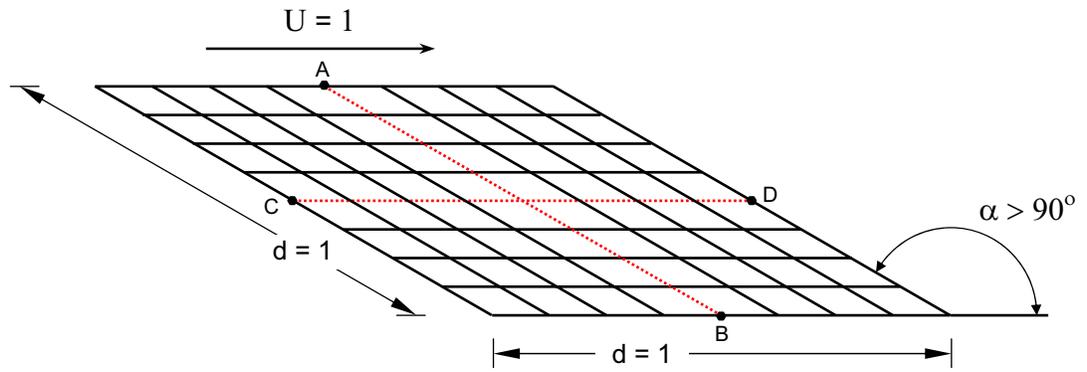

a) skewed cavity with $\alpha > 90°$

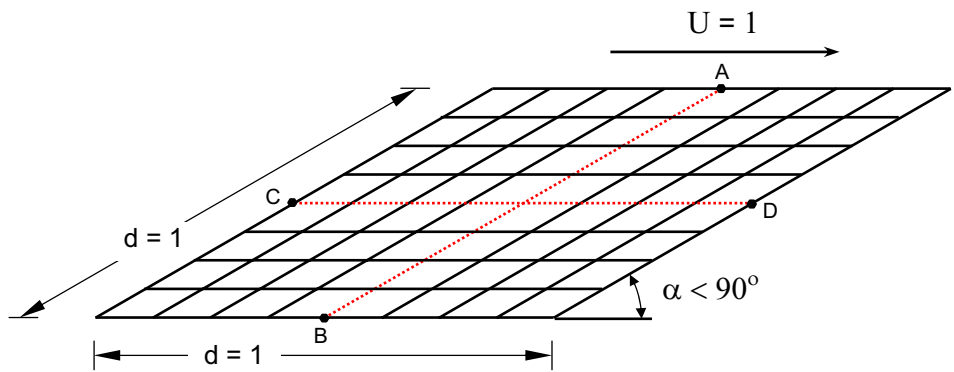

b) skewed cavity with $\alpha < 90°$

Fig. 1. Schematic view of driven skewed cavity flow

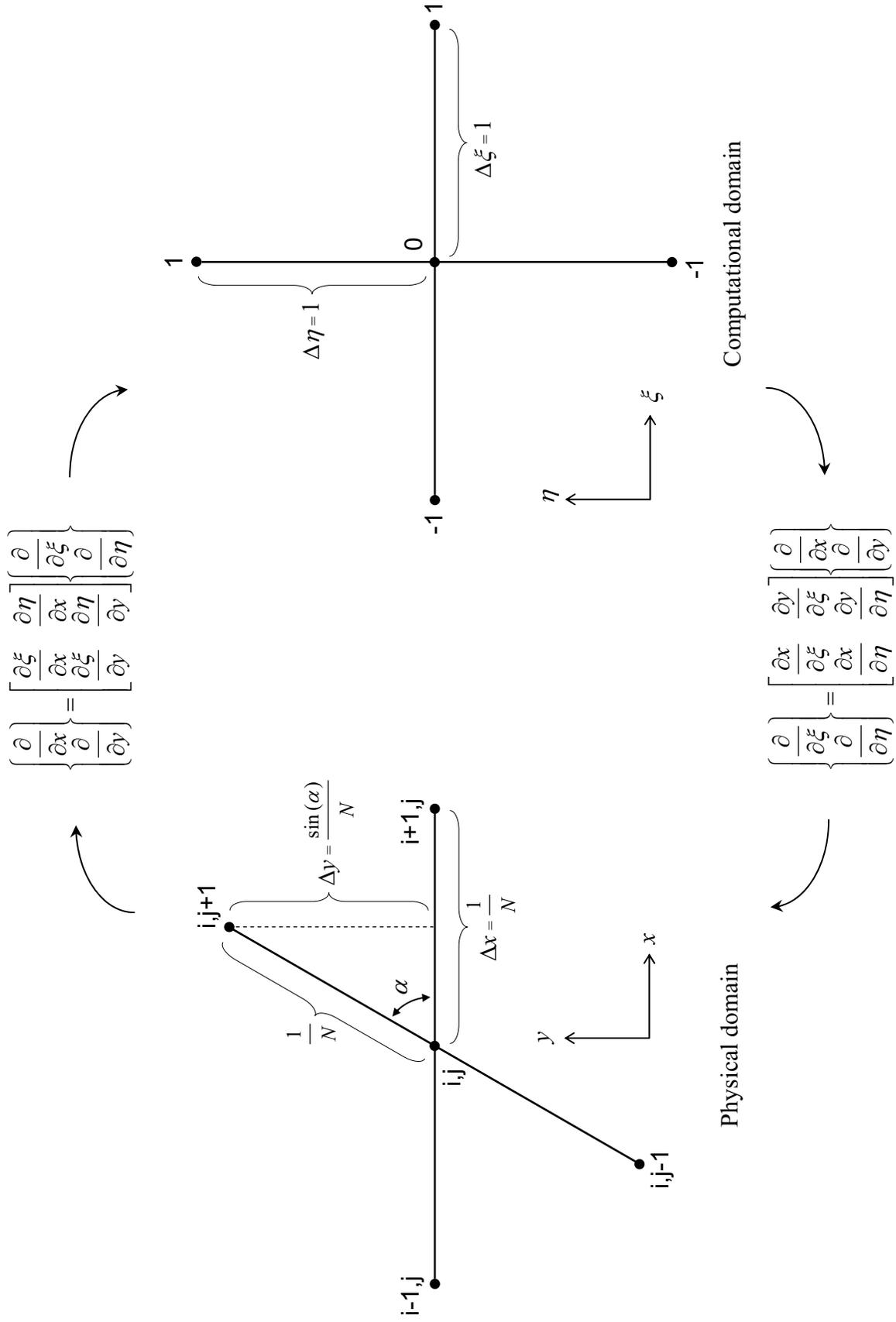

Fig. 2. Transformation of the physical domain to computational domain

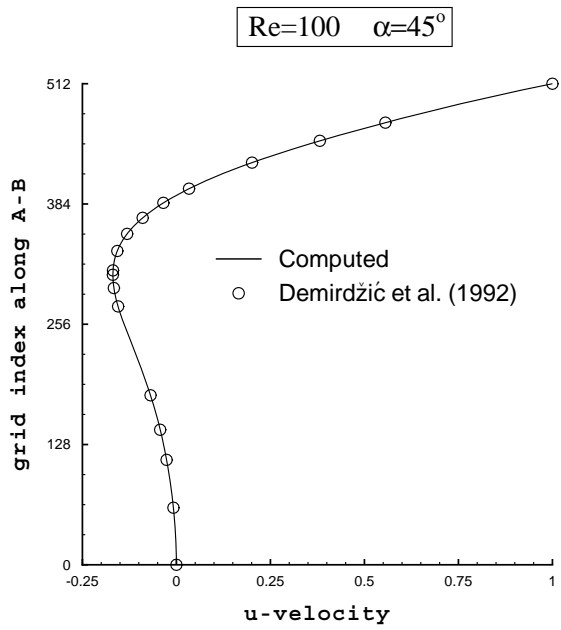
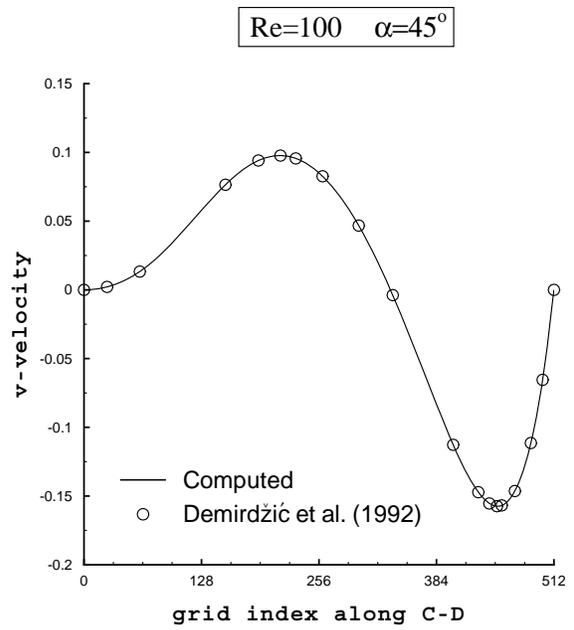
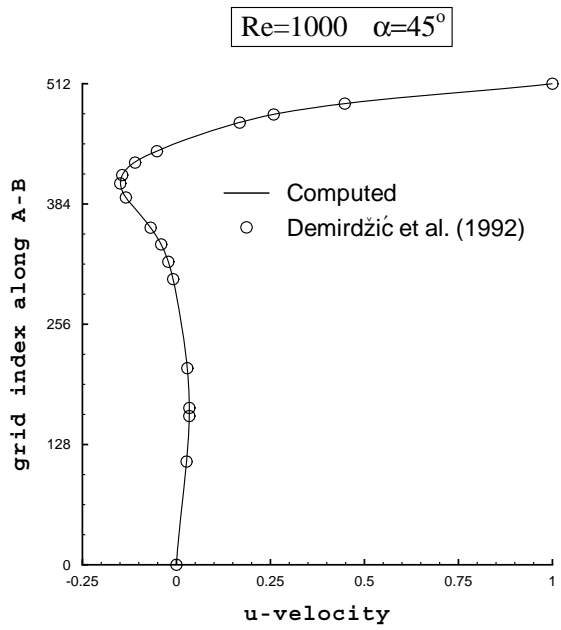
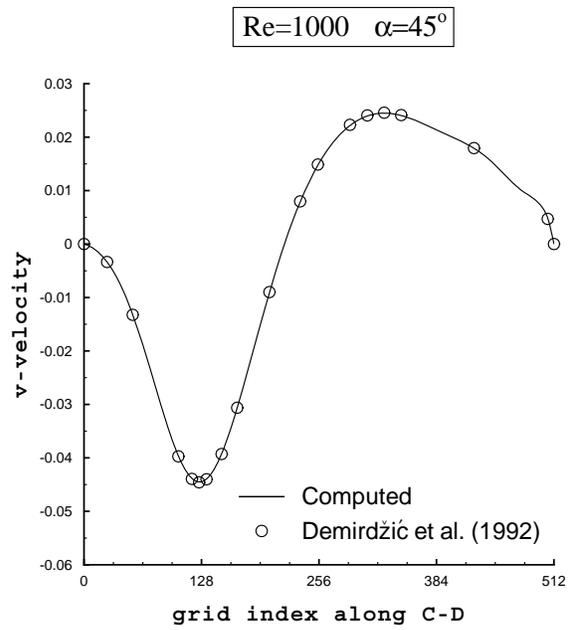

Fig. 3. Comparison of u-velocity along line A-B and v-velocity along line C-D, for Re=100 and 1000, for skew angle α=45°

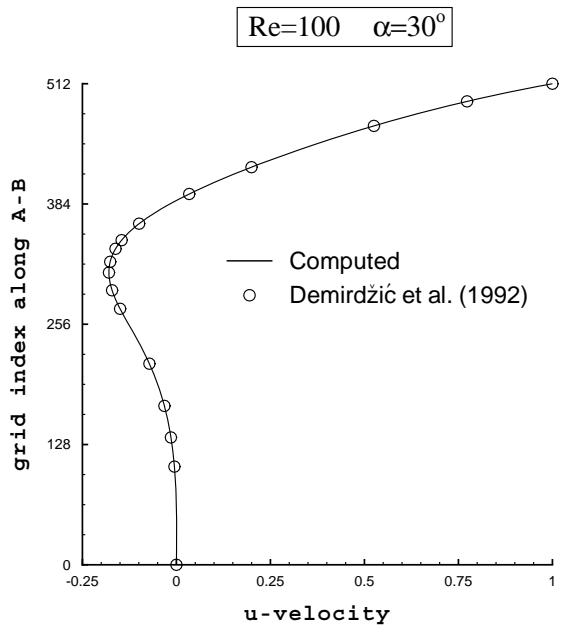
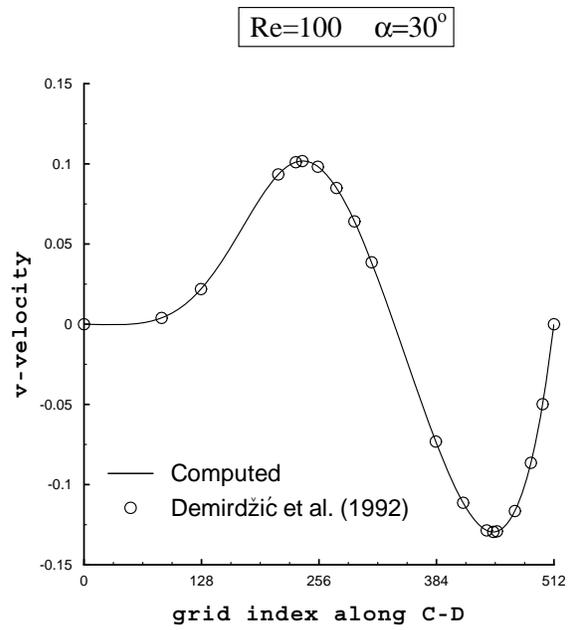
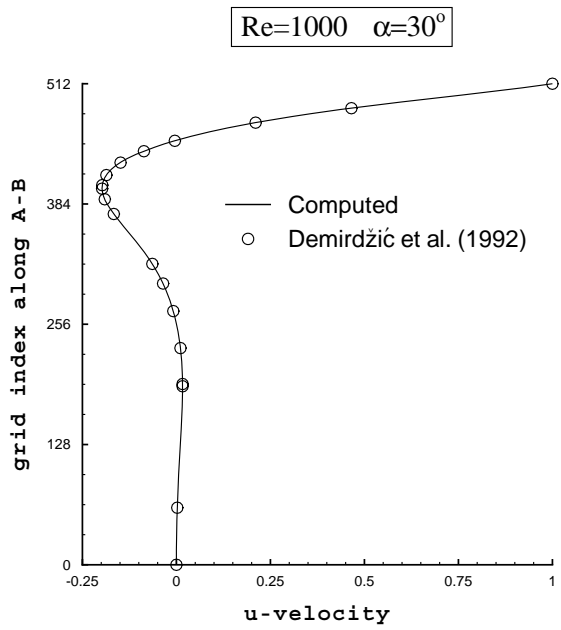
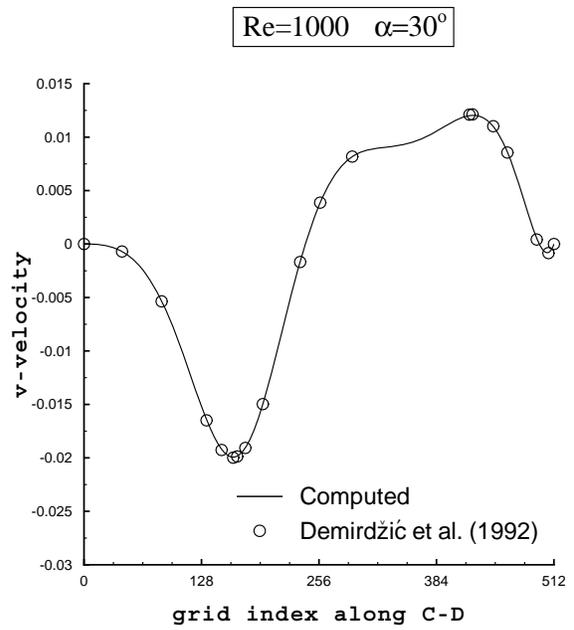

Fig. 4. Comparison of u-velocity along line A-B and v-velocity along line C-D, for Re=100 and 1000, for skew angle α=30°

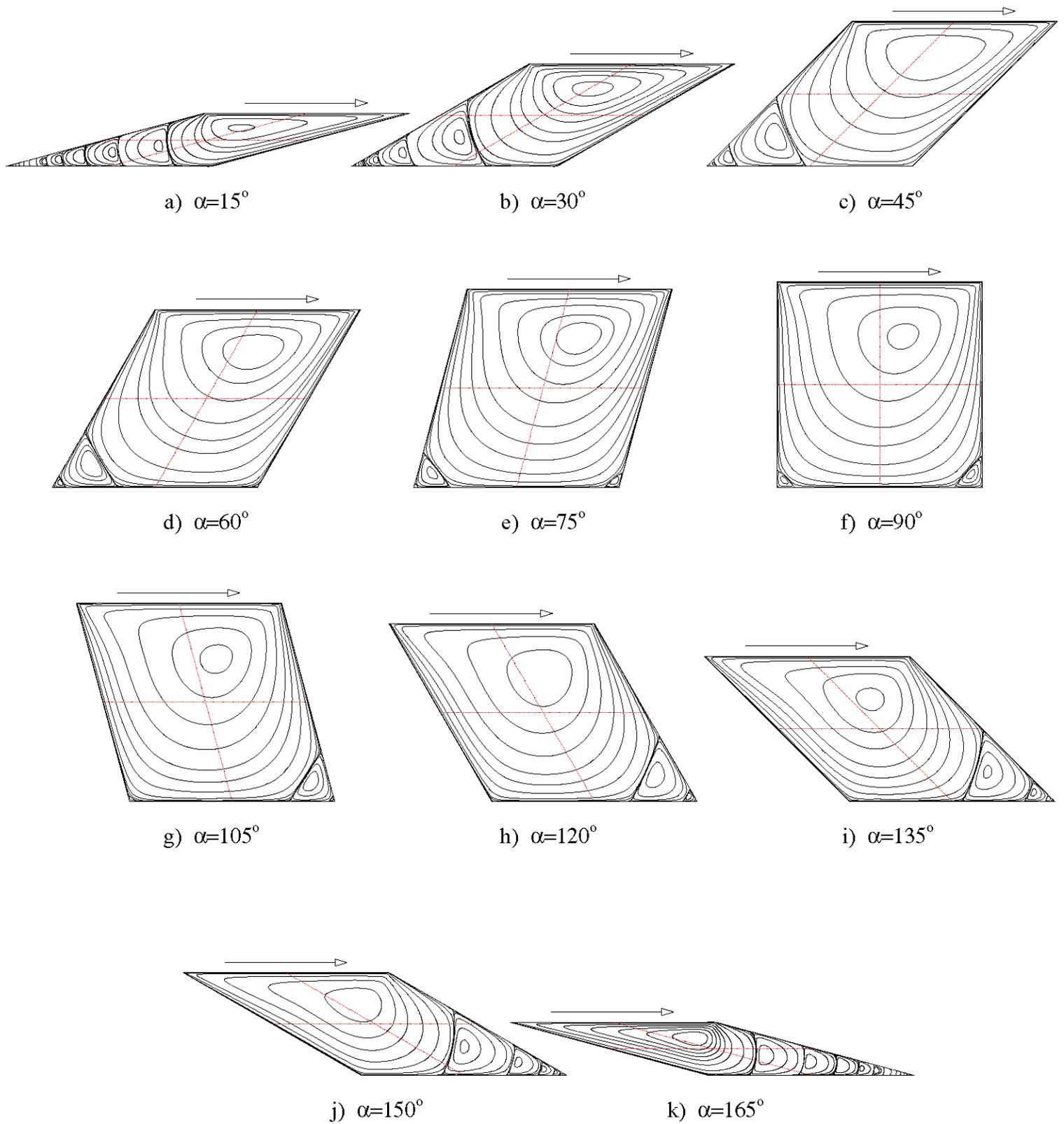

Fig. 5. Streamline contours of driven skewed cavity flow for Re=100 from $\alpha=15°$ to $\alpha=165°$

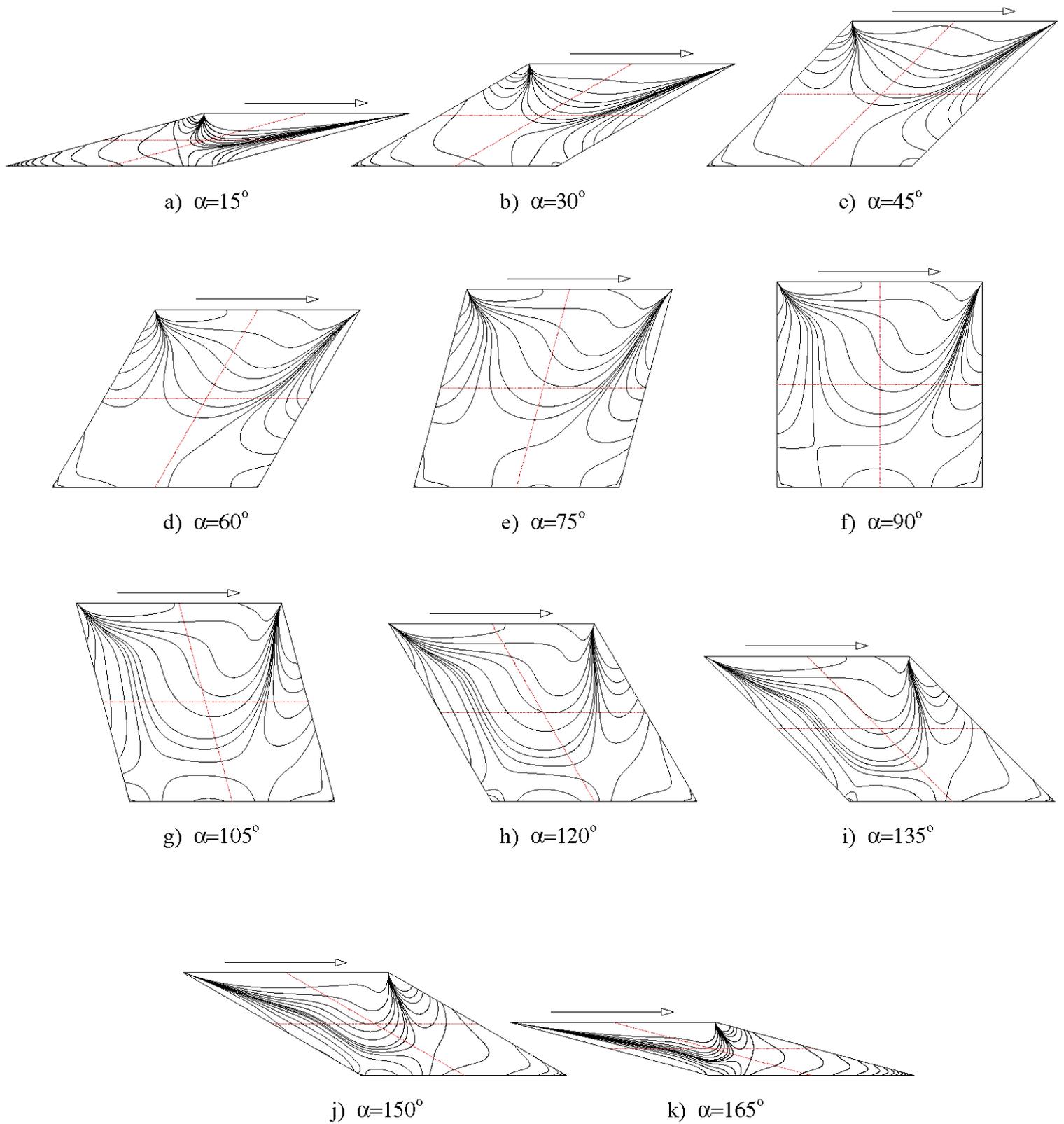

Fig. 6. Vorticity contours of driven skewed cavity flow for Re=100 from $\alpha=15°$ to $\alpha=165°$

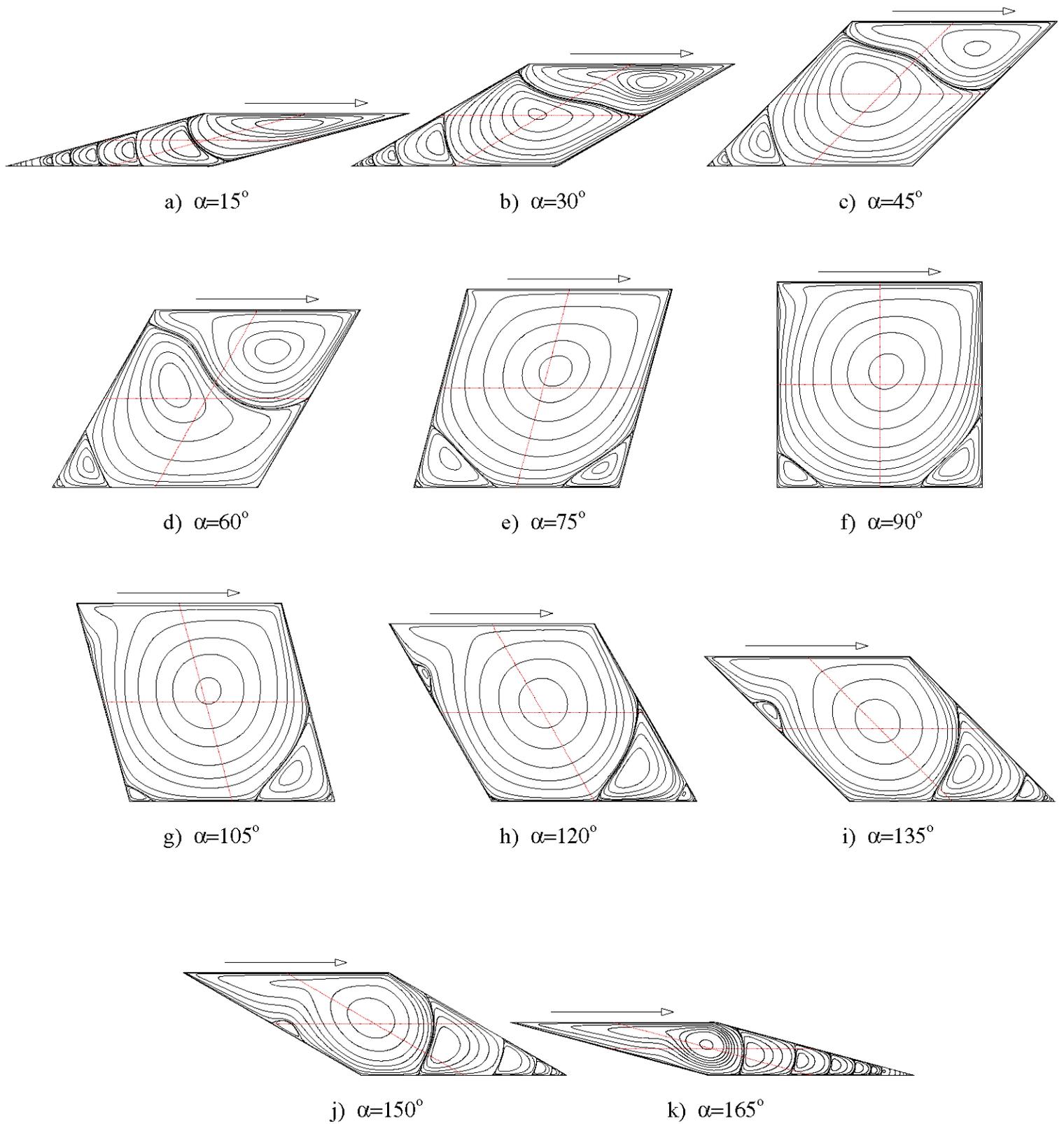

Fig. 7. Streamline contours of driven skewed cavity flow for Re=1000 from $\alpha=15°$ to $\alpha=165°$

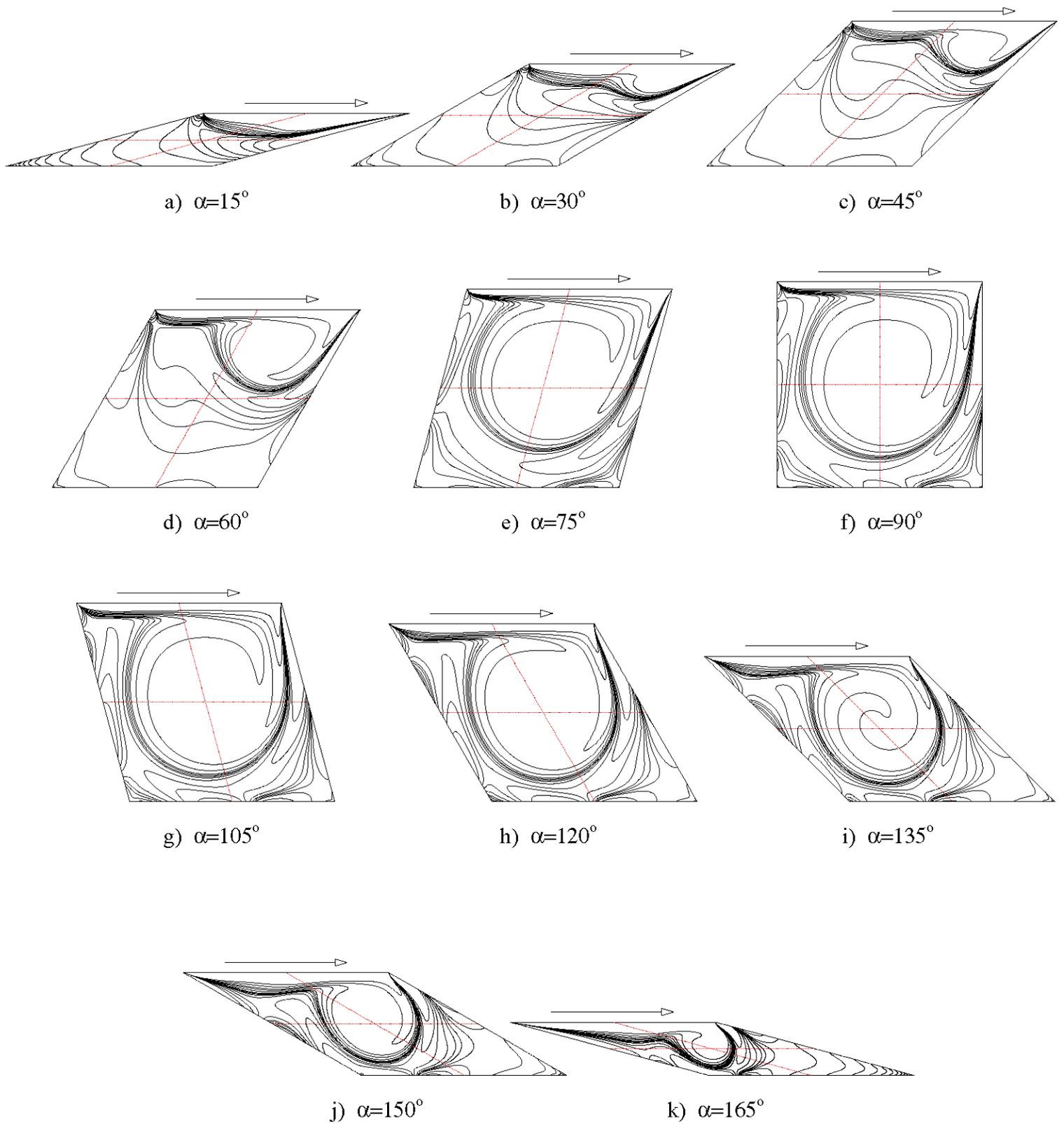

Fig. 8. Vorticity contours of driven skewed cavity flow for Re=1000 from $\alpha=15°$ to $\alpha=165°$

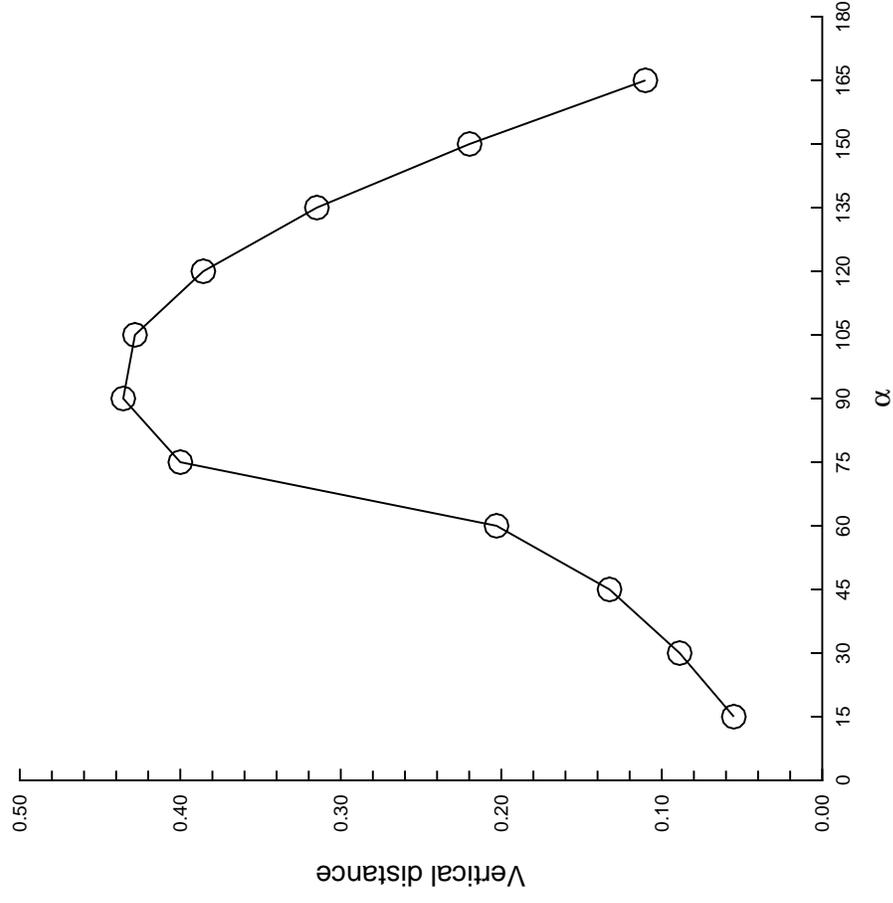

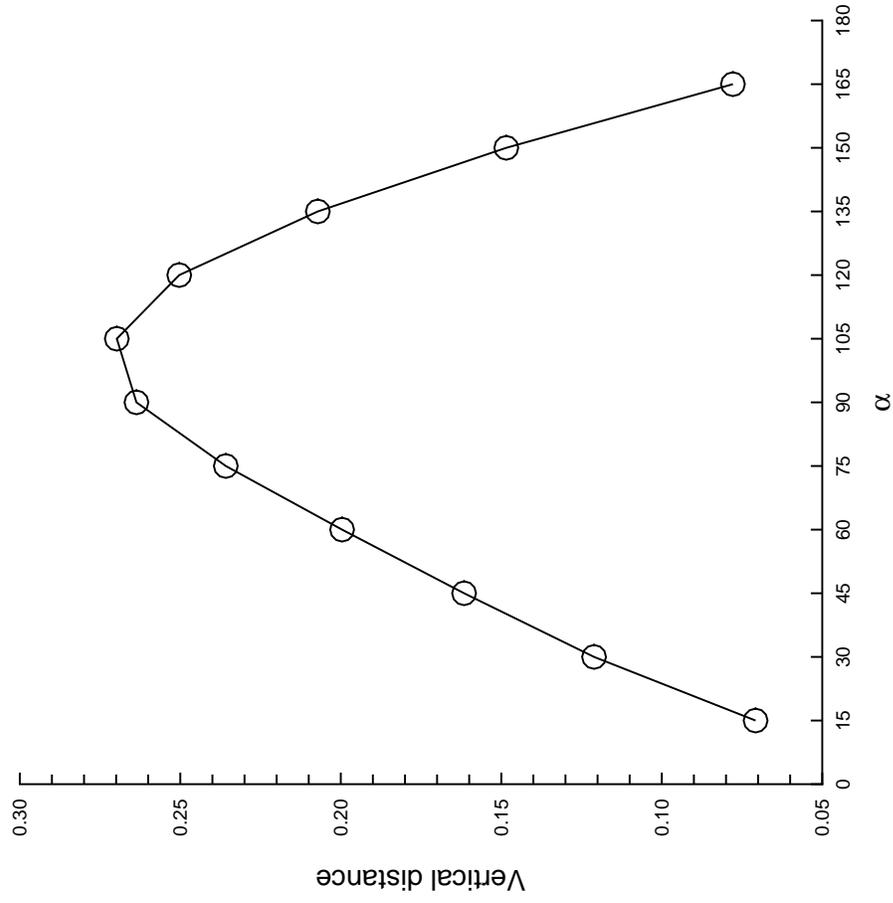

Fig. 9. The vertical distance between the center of the primary vortex and the top moving lid versus the skew angle

| Reference | Grid | Accuracy | $\psi$ | $\omega$ | $x$ | $y$ |
|---|---|---|---|---|---|---|
| Present Study | 513x513 | $\Delta h^2$ | 0.11872 | 2.06476 | 0.5313 | 0.5645 |
| Erturk *et al.* [6] | 513x513 | $\Delta h^2$ | 0.118722 | 2.064765 | 0.5313 | 0.5645 |
| Botella & Peyret [4] | N=160 | N=160 | 0.1189366 | 2.067753 | 0.5308 | 0.5652 |
| Schreiber & Keller [21] | Extrapolated | $\Delta h^6$ | 0.11894 | 2.0677 | - | - |
| Li *et al.* [12] | 129x129 | $\Delta h^4$ | 0.118448 | 2.05876 | 0.5313 | 0.5652 |
| Wright & Gaskell [30] | 1024x1024 | $\Delta h^2$ | 0.118821 | 2.06337 | 0.5313 | 0.5625 |
| Erturk & Gokcol [7] | 601x601 | $\Delta h^4$ | 0.118938 | 2.067760 | 0.5300 | 0.5650 |
| Benjamin & Denny [2] | Extrapolated | high order | 0.1193 | 2.078 | - | - |
| Nishida & Satofuka [16] | 129x129 | $\Delta h^8$ | 0.119004 | 2.068546 | 0.5313 | 0.5625 |

Table 1) Comparison of the properties of the primary vortex for square driven cavity flow; the maximum streamfunction value, the vorticity value and the location of the center, for *Re*=1000

| Skew Angle | | | Re=100 | | Re=1000 | |
|---|---|---|---|---|---|---|
| | | | min | max | min | max |
| α = 30° | Present study (513×513) | $\psi$ (x,y) | -5.3139E-02 (1.1680 , 0.3789) | 5.5343E-05 (0.5262 , 0.1426) | -3.8544E-02 (1.4562 , 0.4111) | 4.1358E-03 (0.9024 , 0.2549) |
| | Demirdžić et al. [5] (320×320) | $\psi$ (x,y) | -5.3135E-02 (1.1664 , 0.3790) | 5.6058E-05 (0.5269 , 0.1433) | -3.8563E-02 (1.4583 , 0.4109) | 4.1494E-03 (0.9039 , 0.2550) |
| | Oosterlee et al. [17] (256×256) | $\psi$ (x,y) | -5.3149E-02 (1.1680 , 0.3789) | 5.6228E-05 (0.5291 , 0.1426) | -3.8600E-02 (1.4565 , 0.4102) | 4.1657E-03 (0.9036 , 0.2559) |
| | Shklyar and Arbel [22] (320×320) | $\psi$ (x,y) | -5.3004E-02 (1.1674 , 0.3781) | 5.7000E-05 (0.5211 , 0.1543) | -3.8185E-02 (1.4583 , 0.4109) | 3.8891E-03 (0.8901 , 0.2645) |
| | Louaked et al. [13] (120×120) | $\psi$ (x,y) | - - | - - | -3.9000E-02 (1.4540 , 0.4080) | 4.3120E-03 (0.8980 , 0.2560) |
| α = 45° | Present study (513×513) | $\psi$ (x,y) | -7.0232E-02 (1.1119 , 0.5455) | 3.6724E-05 (0.3395 , 0.1422) | -5.3423E-02 (1.3148 , 0.5745) | 1.0024E-02 (0.7780 , 0.3991) |
| | Demirdžić et al. [5] (320×320) | $\psi$ (x,y) | -7.0226E-02 (1.1100 , 0.5464) | 3.6831E-05 (0.3387 , 0.1431) | -5.3507E-02 (1.3130 , 0.5740) | 1.0039E-02 (0.7766 , 0.3985) |
| | Oosterlee et al. [17] (256×256) | $\psi$ (x,y) | -7.0238E-02 (1.1100 , 0.5469) | 3.6932E-05 (0.3390 , 0.1409) | -5.3523E-02 (1.3128 , 0.5745) | 1.0039E-02 (0.7775 , 0.4005) |
| | Shklyar and Arbel [22] (320×320) | $\psi$ (x,y) | -7.0129E-02 (1.1146 , 0.5458) | 3.9227E-05 (0.3208 , 0.1989) | -5.2553E-02 (1.3120 , 0.5745) | 1.0039E-02 (0.7766 , 0.3985) |
| | Louaked et al. [13] (120×120) | $\psi$ (x,y) | - - | - - | -5.4690E-02 (1.3100 , 0.5700) | 1.0170E-02 (0.7760 , 0.3980) |

Table 2) Comparison of the minimum and maximum streamfunction value and the location of these points, for Reynolds number of 100 and 1000, for skew angles of 30° and 45°

|  | $Q_{AB} = \dfrac{\left|\int u\,dy + \int v\,dx\right|}{\int |u|\,dy + \int |v|\,dx}$ | | $Q_{CD} = \dfrac{\left|\int v\,dx\right|}{\int |v|\,dx}$ | |
|---|---|---|---|---|
| Skew Angle | $Re$=100 | $Re$=1000 | $Re$=100 | $Re$=1000 |
| $\alpha$ =15° | 1.999E-05 | 2.275E-05 | 1.699E-05 | 7.794E-05 |
| $\alpha$ =30° | 2.089E-05 | 4.485E-05 | 2.692E-05 | 9.423E-06 |
| $\alpha$ =45° | 2.056E-05 | 5.841E-05 | 2.976E-05 | 1.552E-05 |
| $\alpha$ =60° | 2.054E-05 | 3.660E-05 | 2.225E-05 | 2.112E-06 |
| $\alpha$ =75° | 2.172E-05 | 3.849E-05 | 1.065E-05 | 1.371E-05 |
| $\alpha$ =90° | 2.420E-05 | 5.096E-05 | 2.855E-07 | 5.988E-06 |
| $\alpha$ =105° | 2.444E-05 | 5.546E-05 | 6.075E-06 | 1.017E-05 |
| $\alpha$ =120° | 2.399E-05 | 4.875E-05 | 8.410E-06 | 5.312E-06 |
| $\alpha$ =135° | 2.256E-05 | 4.232E-05 | 8.836E-06 | 1.049E-06 |
| $\alpha$ =150° | 2.058E-05 | 2.963E-05 | 1.011E-05 | 3.649E-07 |
| $\alpha$ =165° | 1.842E-05 | 1.654E-05 | 1.165E-05 | 5.009E-06 |

Table 3) Normalized volumetric flow rates through sections A-B and C-D

|  | | Re=100 | | Re=1000 | |
| --- | --- | --- | --- | --- | --- |
| Skew Angle | | min | max | min | max |
| $\alpha = 15°$ | $\psi$  $\omega$ | -3.1296E-02  -9.79337 | 5.7912E-05  5.6154E-02 | -2.4788E-02  -13.90465 | 1.9689E-04  1.3912E-01 |
|  | $(x, y)$ | (1.1393, 0.1880) | (0.7447, 0.0996) | (1.4009, 0.2037) | (0.8705, 0.1087) |
| $\alpha = 30°$ | $\psi$  $\omega$ | -5.3139E-02  -5.89608 | 5.5343E-05  2.7263E-02 | -3.8544E-02  -9.92219 | 4.1358E-03  5.5961E-01 |
|  | $(x, y)$ | (1.1680, 0.3789) | (0.5262, 0.1426) | (1.4562, 0.4111) | (0.9024, 0.2549) |
| $\alpha = 45°$ | $\psi$  $\omega$ | -7.0232E-02  -4.61452 | 3.6724E-05  1.8247E-02 | -5.3423E-02  -6.95543 | 1.0024E-02  6.2686E-01 |
|  | $(x, y)$ | (1.1119, 0.5455) | (0.3395, 0.1422) | (1.3148, 0.5745) | (0.7780, 0.3991) |
| $\alpha = 60°$ | $\psi$  $\omega$ | -8.4736E-02  -3.91398 | 1.8172E-05  1.5238E-02 | -7.5489E-02  -4.55744 | 1.1329E-02  7.6643E-01 |
|  | $(x, y)$ | (0.9844, 0.6664) | (0.1914, 0.1116) | (1.0703, 0.6631) | (0.5879, 0.5041) |
| $\alpha = 75°$ | $\psi$  $\omega$ | -9.6451E-02  -3.43315 | 6.8985E-06  1.4337E-02 | -1.1315E-01  -2.19235 | 1.1701E-03  1.07451 |
|  | $(x, y)$ | (0.8089, 0.7301) | (0.0890, 0.0698) | (0.6888, 0.5660) | (0.9095, 0.0924) |
| $\alpha = 90°$ | $\psi$  $\omega$ | -1.0351E-01  -3.15952 | 1.2755E-05  3.4981E-02 | -1.1872E-01  -2.06476 | 1.7275E-03  1.11098 |
|  | $(x, y)$ | (0.6152, 0.7363) | (0.9434, 0.0625) | (0.5313, 0.5645) | (0.8633, 0.1113) |
| $\alpha = 105°$ | $\psi$  $\omega$ | -1.0419E-01  -3.14779 | 3.8987E-05  4.6978E-02 | -1.1709E-01  -2.13999 | 2.5815E-03  1.11932 |
|  | $(x, y)$ | (0.4267, 0.6961) | (0.8844, 0.0962) | (0.3852, 0.5377) | (0.7869, 0.1321) |
| $\alpha = 120°$ | $\psi$  $\omega$ | -9.7679E-02  -3.41108 | 7.6889E-05  5.4836E-02 | -1.0862E-01  -2.40556 | 3.3676E-03  1.14674 |
|  | $(x, y)$ | (0.2539, 0.6157) | (0.7949, 0.1252) | (0.2520, 0.4804) | (0.6836, 0.1455) |
| $\alpha = 135°$ | $\psi$  $\omega$ | -8.3704E-02  -4.06484 | 1.0764E-04  5.7419E-02 | -9.3512E-02  -2.99180 | 3.7805E-03  1.19738 |
|  | $(x, y)$ | (0.1055, 0.4999) | (0.6708, 0.1436) | (0.1390, 0.3922) | (0.5554, 0.1478) |
| $\alpha = 150°$ | $\psi$  $\omega$ | -6.2347E-02  -5.46336 | 1.0998E-04  6.0074E-02 | -7.1681E-02  -4.35978 | 3.4432E-03  1.32398 |
|  | $(x, y)$ | (-0.0074, 0.3516) | (0.5003, 0.1396) | (0.0478, 0.2803) | (0.3994, 0.1348) |
| $\alpha = 165°$ | $\psi$  $\omega$ | -3.4257E-02  -9.62563 | 7.9703E-05  8.0827E-02 | -4.1542E-02  -8.90771 | 1.8408E-03  1.60475 |
|  | $(x, y)$ | (-0.0660, 0.1810) | (0.2727, 0.0986) | (-0.0097, 0.1486) | (0.2140, 0.0976) |

Table 4) Tabulated minimum and maximum streamfunction values, the vorticity values and the locations, for Reynolds number of 100 and 1000, for various skew angles

| Grid index | $\alpha = 15°$ | $\alpha = 30°$ | $\alpha = 45°$ | $\alpha = 60°$ | $\alpha = 75°$ | $\alpha = 90°$ | $\alpha = 105°$ | $\alpha = 120°$ | $\alpha = 135°$ | $\alpha = 150°$ | $\alpha = 165°$ |
|---|---|---|---|---|---|---|---|---|---|---|---|
| 0 | 0.0000 | 0.0000 | 0.0000 | 0.0000 | 0.0000 | 0.0000 | 0.0000 | 0.0000 | 0.0000 | 0.0000 | 0.0000 |
| 32 | -1.046E-07 | 5.348E-04 | -2.389E-03 | -1.104E-02 | -2.550E-02 | -4.196E-02 | -4.948E-02 | -3.817E-02 | -1.339E-02 | 8.356E-04 | 5.251E-06 |
| 64 | 8.216E-05 | -1.246E-04 | -8.901E-03 | -2.542E-02 | -5.002E-02 | -7.711E-02 | -9.142E-02 | -7.886E-02 | -3.997E-02 | -3.508E-03 | 1.867E-04 |
| 96 | 4.236E-04 | -3.723E-03 | -1.949E-02 | -4.262E-02 | -7.499E-02 | -1.098E-01 | -1.297E-01 | -1.194E-01 | -7.459E-02 | -1.722E-02 | 7.885E-04 |
| 128 | 6.686E-04 | -1.193E-02 | -3.426E-02 | -6.282E-02 | -1.014E-01 | -1.419E-01 | -1.663E-01 | -1.596E-01 | -1.143E-01 | -4.174E-02 | 7.299E-04 |
| 160 | -1.509E-03 | -2.646E-02 | -5.373E-02 | -8.627E-02 | -1.292E-01 | -1.727E-01 | -1.995E-01 | -1.974E-01 | -1.571E-01 | -7.746E-02 | -4.517E-03 |
| 192 | -1.228E-02 | -4.960E-02 | -7.846E-02 | -1.125E-01 | -1.563E-01 | -1.984E-01 | -2.242E-01 | -2.265E-01 | -1.976E-01 | -1.250E-01 | -2.378E-02 |
| 224 | -4.672E-02 | -8.374E-02 | -1.080E-01 | -1.393E-01 | -1.785E-01 | -2.129E-01 | -2.328E-01 | -2.374E-01 | -2.237E-01 | -1.772E-01 | -7.796E-02 |
| 256 | -1.246E-01 | -1.274E-01 | -1.390E-01 | -1.616E-01 | -1.892E-01 | -2.091E-01 | -2.185E-01 | -2.223E-01 | -2.215E-01 | -2.088E-01 | -1.743E-01 |
| 288 | -1.969E-01 | -1.675E-01 | -1.634E-01 | -1.717E-01 | -1.808E-01 | -1.821E-01 | -1.792E-01 | -1.799E-01 | -1.861E-01 | -1.940E-01 | -2.037E-01 |
| 320 | -1.794E-01 | -1.776E-01 | -1.669E-01 | -1.594E-01 | -1.478E-01 | -1.313E-01 | -1.189E-01 | -1.172E-01 | -1.261E-01 | -1.404E-01 | -1.544E-01 |
| 352 | -9.556E-02 | -1.309E-01 | -1.315E-01 | -1.146E-01 | -8.742E-02 | -6.026E-02 | -4.446E-02 | -4.358E-02 | -5.514E-02 | -7.342E-02 | -9.361E-02 |
| 384 | 1.746E-02 | -1.756E-02 | -3.945E-02 | -2.745E-02 | 1.595E-03 | 2.785E-02 | 3.966E-02 | 3.604E-02 | 2.086E-02 | 5.348E-04 | -1.142E-02 |
| 416 | 1.587E-01 | 1.544E-01 | 1.217E-01 | 1.149E-01 | 1.280E-01 | 1.404E-01 | 1.416E-01 | 1.318E-01 | 1.157E-01 | 1.033E-01 | 1.153E-01 |
| 448 | 3.461E-01 | 3.727E-01 | 3.544E-01 | 3.323E-01 | 3.203E-01 | 3.105E-01 | 2.975E-01 | 2.824E-01 | 2.710E-01 | 2.751E-01 | 3.049E-01 |
| 480 | 6.096E-01 | 6.403E-01 | 6.477E-01 | 6.365E-01 | 6.184E-01 | 5.974E-01 | 5.762E-01 | 5.583E-01 | 5.504E-01 | 5.612E-01 | 5.844E-01 |
| 512 | 1.0000 | 1.0000 | 1.0000 | 1.0000 | 1.0000 | 1.0000 | 1.0000 | 1.0000 | 1.0000 | 1.0000 | 1.0000 |

Table 5)  Tabulated *u*-velocity profiles along line A-B, for various skew angles, for *Re*=100

| Grid index | $\alpha = 15°$ | $\alpha = 30°$ | $\alpha = 45°$ | $\alpha = 60°$ | $\alpha = 75°$ | $\alpha = 90°$ | $\alpha = 105°$ | $\alpha = 120°$ | $\alpha = 135°$ | $\alpha = 150°$ | $\alpha = 165°$ |
|---|---|---|---|---|---|---|---|---|---|---|---|
| 0 | 0.0000 | 0.0000 | 0.0000 | 0.0000 | 0.0000 | 0.0000 | 0.0000 | 0.0000 | 0.0000 | 0.0000 | 0.0000 |
| 32 | -3.288E-06 | 4.793E-04 | 6.609E-03 | 1.207E-02 | -7.301E-02 | -2.015E-01 | -2.088E-01 | -1.096E-01 | 2.413E-02 | 2.430E-02 | -8.002E-05 |
| 64 | -6.741E-06 | 2.474E-03 | 1.515E-02 | 2.177E-02 | -2.081E-01 | -3.468E-01 | -3.555E-01 | -2.777E-01 | -8.895E-02 | 3.998E-02 | 6.432E-04 |
| 96 | 6.860E-05 | 6.023E-03 | 2.356E-02 | 2.909E-02 | -3.379E-01 | -3.837E-01 | -3.778E-01 | -3.587E-01 | -2.480E-01 | -8.784E-03 | 5.942E-03 |
| 128 | 4.193E-04 | 1.044E-02 | 3.055E-02 | 3.328E-02 | -3.556E-01 | -3.187E-01 | -3.052E-01 | -3.034E-01 | -3.020E-01 | -1.236E-01 | 1.571E-02 |
| 160 | 1.085E-03 | 1.456E-02 | 3.417E-02 | 3.259E-02 | -2.873E-01 | -2.454E-01 | -2.351E-01 | -2.308E-01 | -2.341E-01 | -2.283E-01 | 6.163E-03 |
| 192 | 9.249E-04 | 1.627E-02 | 3.235E-02 | 2.539E-02 | -2.145E-01 | -1.835E-01 | -1.754E-01 | -1.718E-01 | -1.702E-01 | -1.796E-01 | -5.770E-02 |
| 224 | -5.788E-03 | 1.247E-02 | 2.477E-02 | 1.051E-02 | -1.499E-01 | -1.233E-01 | -1.160E-01 | -1.139E-01 | -1.148E-01 | -1.182E-01 | -1.395E-01 |
| 256 | -3.399E-02 | 2.305E-04 | 1.327E-02 | -1.699E-02 | -8.557E-02 | -6.204E-02 | -5.571E-02 | -5.474E-02 | -5.813E-02 | -6.683E-02 | -9.094E-02 |
| 288 | -9.154E-02 | -2.354E-02 | -5.257E-04 | -7.328E-02 | -1.941E-02 | 4.851E-04 | 5.684E-03 | 5.516E-03 | 4.957E-05 | -1.385E-02 | -5.420E-02 |
| 320 | -1.504E-01 | -6.401E-02 | -1.957E-02 | -1.600E-01 | 4.928E-02 | 6.508E-02 | 6.892E-02 | 6.739E-02 | 5.952E-02 | 3.851E-02 | -4.750E-02 |
| 352 | -1.674E-01 | -1.235E-01 | -5.675E-02 | -1.886E-01 | 1.221E-01 | 1.333E-01 | 1.354E-01 | 1.320E-01 | 1.198E-01 | 7.867E-02 | -9.372E-02 |
| 384 | -1.055E-01 | -1.842E-01 | -1.206E-01 | -8.809E-02 | 2.016E-01 | 2.075E-01 | 2.069E-01 | 1.992E-01 | 1.736E-01 | 7.965E-02 | -1.391E-01 |
| 416 | 5.665E-02 | -1.843E-01 | -1.428E-01 | 7.657E-02 | 2.883E-01 | 2.879E-01 | 2.814E-01 | 2.606E-01 | 1.981E-01 | 2.251E-02 | -3.318E-02 |
| 448 | 3.175E-01 | -3.197E-02 | -6.035E-03 | 2.582E-01 | 3.700E-01 | 3.618E-01 | 3.420E-01 | 2.907E-01 | 1.670E-01 | -2.538E-02 | 1.799E-01 |
| 480 | 6.442E-01 | 3.562E-01 | 2.664E-01 | 4.276E-01 | 4.510E-01 | 4.220E-01 | 3.788E-01 | 2.980E-01 | 1.938E-01 | 2.774E-01 | 4.910E-01 |
| 512 | 1.0000 | 1.0000 | 1.0000 | 1.0000 | 1.0000 | 1.0000 | 1.0000 | 1.0000 | 1.0000 | 1.0000 | 1.0000 |

Table 6) Tabulated *u*-velocity profiles along line A-B, for various skew angles, for *Re*=1000

| Grid index | $\alpha = 15°$ | $\alpha = 30°$ | $\alpha = 45°$ | $\alpha = 60°$ | $\alpha = 75°$ | $\alpha = 90°$ | $\alpha = 105°$ | $\alpha = 120°$ | $\alpha = 135°$ | $\alpha = 150°$ | $\alpha = 165°$ |
|---|---|---|---|---|---|---|---|---|---|---|---|
| 0 | 0.0000 | 0.0000 | 0.0000 | 0.0000 | 0.0000 | 0.0000 | 0.0000 | 0.0000 | 0.0000 | 0.0000 | 0.0000 |
| 32 | -1.440E-06 | -2.980E-04 | 3.515E-03 | 1.986E-02 | 5.361E-02 | 9.478E-02 | 1.207E-01 | 1.184E-01 | 9.474E-02 | 6.812E-02 | 4.223E-02 |
| 64 | -7.221E-05 | 8.638E-04 | 1.484E-02 | 4.693E-02 | 9.612E-02 | 1.492E-01 | 1.852E-01 | 1.902E-01 | 1.618E-01 | 1.105E-01 | 6.114E-02 |
| 96 | -4.017E-04 | 7.005E-03 | 3.410E-02 | 7.411E-02 | 1.243E-01 | 1.743E-01 | 2.094E-01 | 2.201E-01 | 2.000E-01 | 1.439E-01 | 6.931E-02 |
| 128 | -3.705E-04 | 2.231E-02 | 5.789E-02 | 9.573E-02 | 1.385E-01 | 1.792E-01 | 2.072E-01 | 2.175E-01 | 2.061E-01 | 1.628E-01 | 7.719E-02 |
| 160 | 5.307E-03 | 4.818E-02 | 8.020E-02 | 1.084E-01 | 1.402E-01 | 1.691E-01 | 1.867E-01 | 1.908E-01 | 1.809E-01 | 1.522E-01 | 8.685E-02 |
| 192 | 3.201E-02 | 7.850E-02 | 9.469E-02 | 1.104E-01 | 1.302E-01 | 1.457E-01 | 1.506E-01 | 1.443E-01 | 1.276E-01 | 1.016E-01 | 7.282E-02 |
| 224 | 9.001E-02 | 9.932E-02 | 9.693E-02 | 1.010E-01 | 1.080E-01 | 1.088E-01 | 9.910E-02 | 7.941E-02 | 4.916E-02 | 1.030E-02 | -1.324E-02 |
| 256 | 1.222E-01 | 9.779E-02 | 8.505E-02 | 7.962E-02 | 7.305E-02 | 5.753E-02 | 3.203E-02 | -2.908E-03 | -4.939E-02 | -1.061E-01 | -1.537E-01 |
| 288 | 7.313E-02 | 7.193E-02 | 5.911E-02 | 4.567E-02 | 2.468E-02 | -7.743E-03 | -4.894E-02 | -9.757E-02 | -1.532E-01 | -1.999E-01 | -1.667E-01 |
| 320 | 1.657E-03 | 2.857E-02 | 2.026E-02 | -5.948E-04 | -3.599E-02 | -8.404E-02 | -1.374E-01 | -1.905E-01 | -2.304E-01 | -2.106E-01 | -6.811E-02 |
| 352 | -4.738E-02 | -2.320E-02 | -2.898E-02 | -5.685E-02 | -1.045E-01 | -1.630E-01 | -2.182E-01 | -2.552E-01 | -2.440E-01 | -1.449E-01 | -1.343E-02 |
| 384 | -7.167E-02 | -7.428E-02 | -8.326E-02 | -1.166E-01 | -1.702E-01 | -2.278E-01 | -2.665E-01 | -2.622E-01 | -1.903E-01 | -6.954E-02 | -1.555E-04 |
| 416 | -7.836E-02 | -1.142E-01 | -1.322E-01 | -1.667E-01 | -2.146E-01 | -2.537E-01 | -2.568E-01 | -2.056E-01 | -1.093E-01 | -2.295E-02 | 6.897E-04 |
| 448 | -7.150E-02 | -1.294E-01 | -1.572E-01 | -1.848E-01 | -2.121E-01 | -2.186E-01 | -1.858E-01 | -1.170E-01 | -4.378E-02 | -3.811E-03 | 1.537E-04 |
| 480 | -4.890E-02 | -1.006E-01 | -1.284E-01 | -1.406E-01 | -1.415E-01 | -1.233E-01 | -8.469E-02 | -4.007E-02 | -9.214E-03 | 5.144E-04 | 5.610E-06 |
| 512 | 0.0000 | 0.0000 | 0.0000 | 0.0000 | 0.0000 | 0.0000 | 0.0000 | 0.0000 | 0.0000 | 0.0000 | 0.0000 |

Table 7) Tabulated *v*-velocity profiles along line C-D, for various skew angles, for *Re*=100

| Grid index | $\alpha = 15°$ | $\alpha = 30°$ | $\alpha = 45°$ | $\alpha = 60°$ | $\alpha = 75°$ | $\alpha = 90°$ | $\alpha = 105°$ | $\alpha = 120°$ | $\alpha = 135°$ | $\alpha = 150°$ | $\alpha = 165°$ |
|---|---|---|---|---|---|---|---|---|---|---|---|
| 0 | 0.0000 | 0.0000 | 0.0000 | 0.0000 | 0.0000 | 0.0000 | 0.0000 | 0.0000 | 0.0000 | 0.0000 | 0.0000 |
| 32 | 1.531E-06 | -3.406E-04 | -5.132E-03 | -1.717E-02 | 1.809E-01 | 2.799E-01 | 2.678E-01 | 1.601E-01 | 3.646E-02 | -8.451E-04 | 2.827E-02 |
| 64 | 1.500E-07 | -2.408E-03 | -1.882E-02 | -4.224E-02 | 2.846E-01 | 3.641E-01 | 3.776E-01 | 3.384E-01 | 1.954E-01 | 3.264E-02 | 2.674E-02 |
| 96 | -1.003E-04 | -7.569E-03 | -3.661E-02 | -5.437E-02 | 3.428E-01 | 3.672E-01 | 3.778E-01 | 3.821E-01 | 3.449E-01 | 1.562E-01 | 1.232E-02 |
| 128 | -6.255E-04 | -1.518E-02 | -4.446E-02 | -4.211E-02 | 3.211E-01 | 3.067E-01 | 3.119E-01 | 3.301E-01 | 3.545E-01 | 3.107E-01 | 4.428E-02 |
| 160 | -1.657E-03 | -1.990E-02 | -3.446E-02 | -1.662E-02 | 2.530E-01 | 2.310E-01 | 2.317E-01 | 2.450E-01 | 2.757E-01 | 3.177E-01 | 1.687E-01 |
| 192 | -3.826E-04 | -1.574E-02 | -1.513E-02 | 7.834E-03 | 1.809E-01 | 1.604E-01 | 1.571E-01 | 1.624E-01 | 1.790E-01 | 2.164E-01 | 2.433E-01 |
| 224 | 9.904E-03 | -5.336E-03 | 2.777E-03 | 2.583E-02 | 1.134E-01 | 9.287E-02 | 8.576E-02 | 8.412E-02 | 8.705E-02 | 9.592E-02 | 1.261E-01 |
| 256 | 2.484E-02 | 3.619E-03 | 1.512E-02 | 3.963E-02 | 4.721E-02 | 2.581E-02 | 1.493E-02 | 6.477E-03 | -3.297E-03 | -2.199E-02 | -6.893E-02 |
| 288 | 2.973E-02 | 7.896E-03 | 2.195E-02 | 5.622E-02 | -1.930E-02 | -4.173E-02 | -5.646E-02 | -7.190E-02 | -9.466E-02 | -1.403E-01 | -3.282E-01 |
| 320 | 2.448E-02 | 8.976E-03 | 2.441E-02 | 7.605E-02 | -8.676E-02 | -1.106E-01 | -1.293E-01 | -1.514E-01 | -1.856E-01 | -2.676E-01 | -2.648E-01 |
| 352 | 1.505E-02 | 9.378E-03 | 2.372E-02 | 7.994E-02 | -1.557E-01 | -1.814E-01 | -2.034E-01 | -2.300E-01 | -2.762E-01 | -4.399E-01 | -9.856E-03 |
| 384 | 3.636E-03 | 1.056E-02 | 2.136E-02 | 4.498E-02 | -2.259E-01 | -2.529E-01 | -2.766E-01 | -3.096E-01 | -4.084E-01 | -2.632E-01 | 1.607E-02 |
| 416 | -1.236E-02 | 1.194E-02 | 1.879E-02 | -2.239E-02 | -3.025E-01 | -3.311E-01 | -3.647E-01 | -4.350E-01 | -4.205E-01 | -2.209E-02 | 6.206E-03 |
| 448 | -3.536E-02 | 1.075E-02 | 1.475E-02 | -7.703E-02 | -4.293E-01 | -4.672E-01 | -5.040E-01 | -4.339E-01 | -9.916E-02 | 1.664E-02 | 7.531E-04 |
| 480 | -4.864E-02 | 3.899E-03 | 9.615E-03 | -5.713E-02 | -4.796E-01 | -4.548E-01 | -3.097E-01 | -8.699E-02 | 1.292E-02 | 1.051E-02 | -2.054E-05 |
| 512 | 0.0000 | 0.0000 | 0.0000 | 0.0000 | 0.0000 | 0.0000 | 0.0000 | 0.0000 | 0.0000 | 0.0000 | 0.0000 |

Table 8) Tabulated *v*-velocity profiles along line C-D, for various skew angles, for *Re*=1000